\documentclass{PoS}

\title{Fast aperiodic variability in the black hole binary GRS 1915+105: the
timing signature of relativistic ejection events}

\ShortTitle{The timing signature of relativistic ejection events: GRS 1915+105}

\author{\speaker{Paolo Soleri}\\
        Astronomical Institute ``A. Pannekoek", University of Amsterdam and
	Center for High-Energy Astrophysics, Kruislaan 403, NL-1098 SJ,
	Amsterdam, The Netherlands\\
        E-mail: \email{psoleri@science.uva.nl}}

\author{{Tomaso Belloni}\\
        INAF-Brera Astronomical Observatory, Via E. Bianchi 46, I-23807,
        Merate, Italy\\
        E-mail: \email{tomaso.belloni@brera.inaf.it}}

\author{{Piergiorgio Casella}\\
        Astronomical Institute ``A. Pannekoek", University of Amsterdam and
	Center for High-Energy Astrophysics, Kruislaan 403, NL-1098 SJ,
	Amsterdam, The Netherlands\\
        E-mail: \email{casella@science.uva.nl}}

\abstract{We present X-ray observations of the Black-Hole Binary GRS 1915+105 made
with the RXTE (Rossi X-ray Timing Explorer). 
We concentrated on timing analysis of the strong variability
focusing on its aperiodic variability on short ($<$1s) time
scales. In the power density
spectra, we found a feature which is seen in many transient systems, but until now
was not detected in GRS 1915+105 due to its elusiveness. As this feature has been
associated to the collimation and emission of superluminal relativistic jets visible
in the radio band, its presence on the prototypical galactic jet source strengthens
this connection.
Since its discovery in 1992, GRS 1915+105 was considered a peculiar source.
Our results suggest that its general behaviour is similar to that of other
black-hole binaries, but its
state-transitions are much faster and difficult to analyze.

}

\FullConference{VI Microquasar Workshop: Microquasars and Beyond\\
		September 18-22 2006\\
		Societ\`a del Casino, Como, Italy}

\begin{document}

\section{Introduction} \label{par:intro}
Systematic variations in the energy spectra and intensity of transient Black Hole Candidates
(BHCs) have
been recently identified in terms of the pattern described in an X-ray Hardness-intensity diagram
(HID, see \cite{2005A&A...440..207B}, \cite{2005ApJ...624..295H}). Four main bright states
(in addition to the quiescent state) have been
found to correspond to different branches/areas of a square-like HID pattern. In this framework
much importance is given to the intermediate states (called Hard intermediate state, HIMS, and
Soft intermediate State, SIMS), and to the transitions between them, identified from the
behaviour in several bands of the electromagnetic spectrum (from radio to hard X-ray, see
\cite{2004MNRAS.355.1105F}) and from the timing properties of the X-ray light curve.\\
Low-frequency Quasi-Periodic Oscillations (LFQPOs) with frequencies ranging from a few mHz to
tens of Hz are a common feature in black hole candidates and have been observed in many galactic
BHCs since the '70s (see \cite{2006csxs.book...39V}, \cite{2006csxs.book..157M} and references
therein). Three main
types of LFQPOs, dubbed Type-A, -B and -C respectively, were originally identified in the light
curve of XTE J1550-564 (\cite{1999ApJ...526L..33W}, \cite{2002ApJ...564..962R}) and are seen in
several sources (see \cite{2005ApJ...629..403C} and references therein). In the context of the
state classification outlined above, it is possible to ascribe the three
LFQPOs to different spectral conditions (see  \cite{2005A&A...440..207B},
\cite{2005ApJ...624..295H}).
The type-C QPO is associated to the (radio loud) HIMS and to the low/hard state. It is a
common QPO seen in almost all BHCs with a variable centroid frequency correlated with the count
rate, a high fractional variability and a high coherence ($Q\, =\, \nu / FWHM \sim 10$).
The type-B QPO has been seen in few systems (see \cite{2005ApJ...629..403C}); it is a transient
oscillation associated to the spectral
transition from the (radio loud) HIMS to the (radio quiet) SIMS and its features are a $\sim$
constant centroid frequency, lower fractional variability and $Q$ than the type-C. Some authors
(\cite{2004MNRAS.355.1105F}, \cite{2004A&A...426..587C}) suggested that these spectral
transitions are in turn associated to the emission and collimation of relativistic radio jets.
These jets are seen in a number of sources (GRS 1915+105, XTE J1550-564, GX 339-4, XTE J1859+226,
etc.) but not in all of
these sources we could resolve the spectral transition to see the transient QPO.
The spectral properties connected to type-A QPO are similar to those introduced for the type-B.
This QPO has been seen in few systems (see \cite{2005ApJ...629..403C}), and is broader, weaker
and less coherent than the type-B QPO.\\
GRS 1915+105 was discovered in 1992 with the GRANAT/WATCH (\cite{1992IAUC.5590....2C},
\cite{1994ApJS...92..469C}). It is the
first galactc source observed to have apparently superluminal relativistc radio jets
(\cite{1994Natur.371...46M}), interpreted as ejection of ultrarelativistc plasma.
 Its radio
variability was discovered to correlate to the hard X-ray flux (\cite{1994A&A...282L..17M}).\\
The rich phenomenology of this source was described (see \cite{2000A&A...355..271B}) in terms of
spectral transitions between three basic states, A, B and C (not to be confused with the names of
the LFQPOs), that give rise to more than a dozen 
variability classes. The non-standard behaviour of GRS 1915+105
was interpreted as that of a source that spends all its time in the Intermediate States, never
reaching the LS or the quiescence (\cite{2004ARA&A..42..317F}). Until now all LFQPOs observed
from this system can be classified as type-C QPOs; altough GRS
1915+105 makes a large number of fast state transitions,which have been positively associated to
radio activity and jet ejection, no type-B QPO has been observed to date.\\
Here we present the discovery with RXTE of the type-B QPO in the X-ray light curve of GRS
1915+105. The QPO was present during fast spectral transitions that can be identified with the HIMS to
SIMS transition observed in other BHCs. 
A complete paper with a detailed description of the data analysis and the results is in preparation.

\section{Observations and data analysis} \label{par:data}
\begin{figure}
\begin{tabular}{c}
\hspace{0.75cm}
\includegraphics[width=12cm]{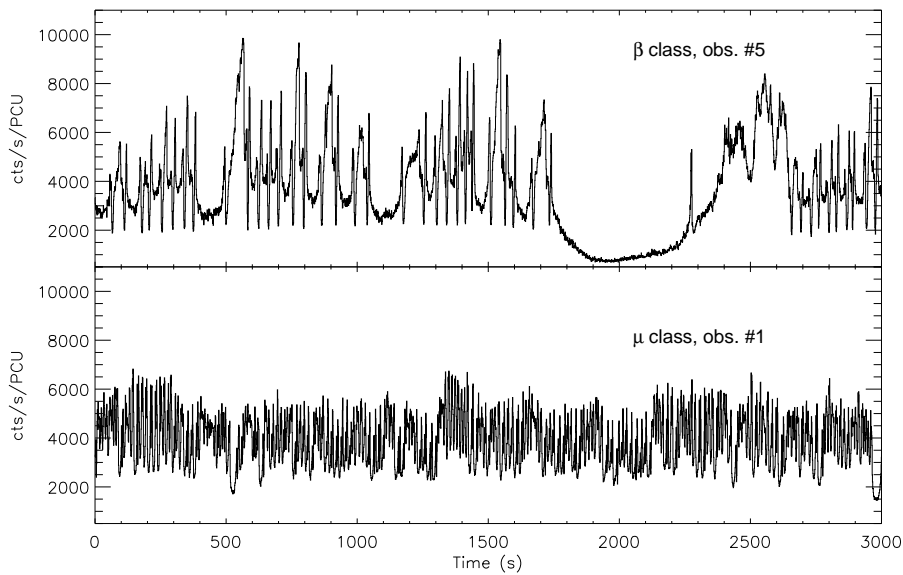}
\end{tabular}
\caption{2-13 keV light curves for the two variability classes examined here. Top panel: observation
from 1997 Dec 17; bottom panel: observation from 1996 Sep 09. 
Bin size is 1 second.}
\label{fig:mu-beta}
\end{figure}

We analyzed five RXTE/PCA observations collected between 1996 September 22 and 1997 December 17. We
chose observations belonging to variability classes $\mu$ and $\beta$ (according to the
classification presented in \cite{2000A&A...355..271B}, see Fig. \ref{fig:mu-beta} for two
examples of light curves) because they correspond to intervals when frequent fast transitions
among the three spectral states are observed. The transitions being fast, the type-B QPO would
not be detected for a single transition, but adding a large number of data corresponding to the
same transition, the signal could become detectable. 
For each observation we produced light curves in the 2-13 keV energy range and two hardness
ratios (HR1 (5-13 keV)$/$(2-13 keV); HR2 (13-38 keV)$/$(2-5 keV)); the bin size was 2 seconds for
all these curves. Given the high count rates, we did not subtract the background.
In addition, for each 2-second time interval, we accumulated a power
density spectrum (PDS) in the 2-5 keV, 5-13 keV and 13-38 keV energy ranges with a time resolution
of $1/128$ sec. We also created 3-dimensional hardness-hardness-intensity diagram, where the
source follows a regular path and we averaged the PDS corresponding to different regions in this
diagram.\\
For a detailed description of all the data analysis see Soleri, Belloni, Casella, to be
submitted.

\section{Results} \label{par:results}

\begin{figure}
\begin{tabular}{c}
\hspace{2.2cm}
\resizebox{10cm}{!}{\includegraphics{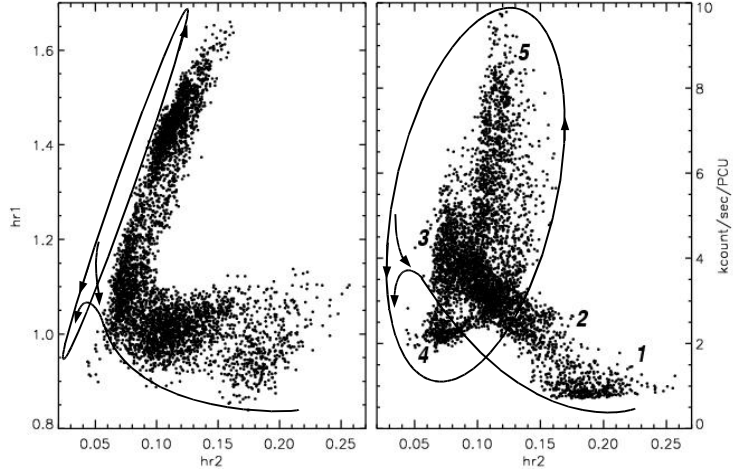}}
\end{tabular}
\caption{Two-dimensional projections of the hardness-hardness-intensity diagram (HHID)
for one analyzed observation. Left side panels: color-color diagram; right side panels:
hardness-intensity diagram. Definitions for the colors can be found in
\S2, the count rate is the 2-13 keV energy band. Arrows indicate the source
regular and repetitive movement. Numbers correspond to the regions discussed in the  
text (see Fig. 3).}
\label{fig:cd_hid_arrows}
\end{figure}

\begin{figure}
\begin{tabular}{c}
\resizebox{14cm}{!}{\includegraphics{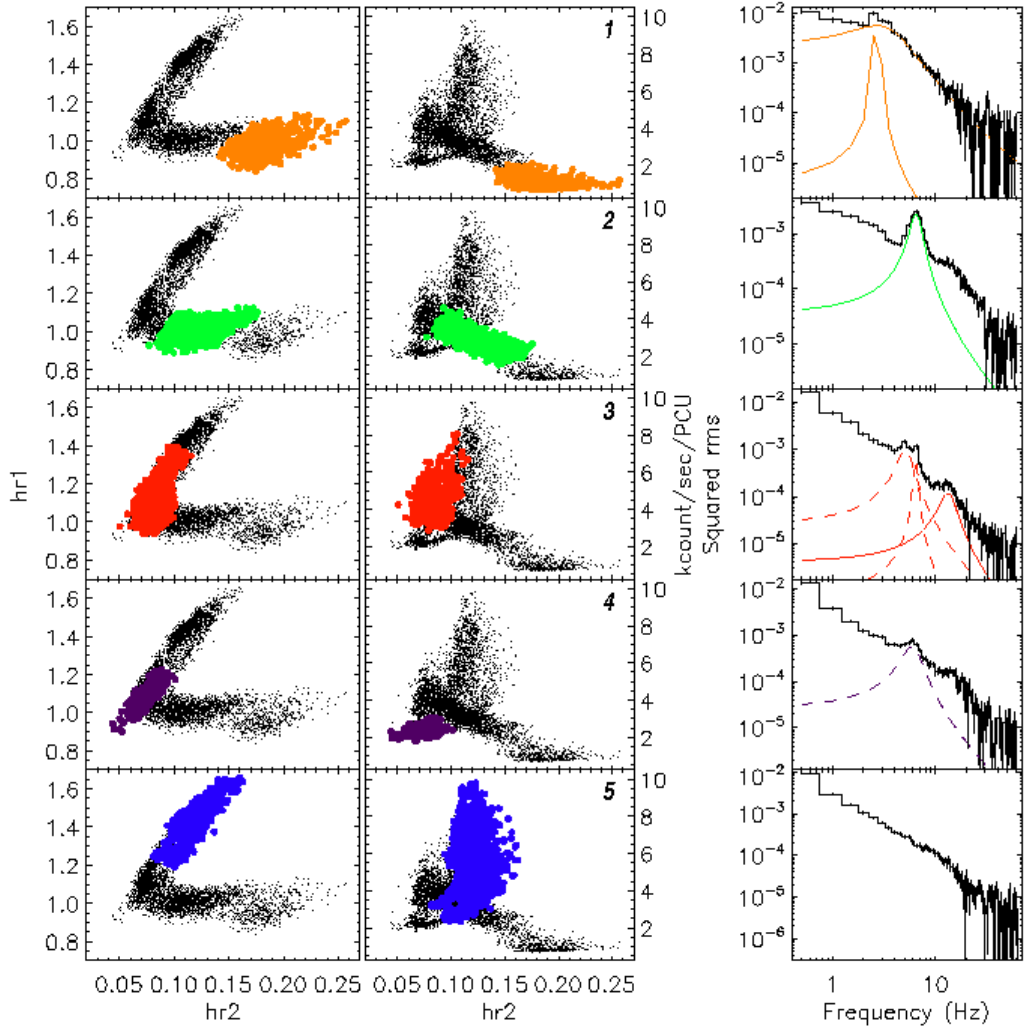}}
\end{tabular}
\caption{Two-dimensional projections of the hardness-hardness-intensity diagram
for observation \#5 and typical power density spectra for each of the five regions.
Left side panel: color-color diagrams; central panels: hardness-intensity diagrams. 
Definitions for hardness ratios can be found in \S2, the count rate is the 2-13 keV
energy band. Colored circles indicate the points corresponding to each region.
Right side panels: power density spectra averaged through the points in the region
in the 5-13 keV energy band. With
dashed lines we plot the Lorentzians used to fit type-B QPOs, with continuous lines the
Lorentzians used to fit type-C QPOs.}
\label{fig:cd_hid_pds}
\end{figure}

An example of 3D HHID diagram is shown by the two-dimensional projections in Fig.
\ref{fig:cd_hid_arrows}. The regular path followed by the source is marked by arrows. Starting
from the region of the HHID with highest HR2 (right side of both the diagrams) we averaged
power density spectra
on small number of points ($\sim$ 300), following the source path, for all the observations.
We found that in this way
it is possible to identify five average behaviours of the PDS, independently from the analyzed
observation. These five behaviours correspond to five separated regions of the source path in the
HHID that are shown in the left and central panels of Figure \ref{fig:cd_hid_pds}.

We describe the average PDS behaviour of these macro-regions in Tab. \ref{tab:zone}: from
this table it is clear that we can use the same macro-region classification for different
observations.
\begin{table}
\centering
\begin{tabular}{c c c c c c}
\hline
\hline
            & \multicolumn{5}{c}{Regions in figure \ref{fig:cd_hid_pds}}\\

Observation N$^{\circ}$ & 1 & 2 & 3 & 4 & 5\\
\hline
1 & - & C & C, B & B & No QPOs \\
2 & Bump - C & C & C, B & No QPOs & No QPOs \\ 
3 & Bump - C & C & C, B & No QPOs & No QPOs \\
4 & - & C & C, B & B & No QPOs \\
5 & Bump - C & C & C, B  & B & No QPOs \\ 
\hline
\end{tabular}
\caption{QPO types detected in the five regions for all considered observations.
Region 1 is not present in observations \# 2 and 3.} \label{tab:zone}
\end{table}
As described in \S \ref{par:data}, we considered observations 
belonging to variability classes $\mu$ and $\beta$. These observations have very similar
light curves and HHID, the only difference being that in class $\beta$ there are long hard
intervals (the smooth stretch in the top panel of Fig. \ref{fig:mu-beta}), which correspond
to region {\em 1} (Orange) and to the hardest part of region {\em 2} (Green)
in the HHID. Since we do not concentrate on the harder intervals, for our
purposes these two classes can be treated as one.\\
We now describe in detail the power density spectra behaviour in all the identified
regions:
\begin{enumerate}
   \item starting from region {\em 1}, we  see a peak, increasing in frequency
   and in quality factor $Q$ (from a $Q<2$ for a centroid frequency $\sim$2 Hz to a $Q>2$
   for higher centroid frequency) and decreasing in fractional {\it rms}
   amplitude moving towards region {\em 2}.
   Its fractional {\it rms} is higher in the hardest band (13-38 keV);
   \item moving within region {\em 2} towards region {\em 3}, the peak continues
   this trend: increasing centroid frequency  and decreasing fractional {\it rms} with the
   same energy dependence;
   \item within region {\em 3}, the  peak still
   increases in frequency and decreases in fractional {\it rms}, to disappear when
   approaching region {\em 4}, corresponding to a frequency of $\sim$15 Hz.
   At variance with what observed in previous regions, 
   an additional peak appears. Its properties are markedly different from those
   of the QPO described above: it shows a much less variable frequency
   (from 4 to 6 Hz) and a fractional {\it rms} independent of its centroid frequency.  
   Its energy dependence is the same as that observed for the other peak.
   Averaging power spectra corresponding only to points close to the boundary to 
   region {\em 4}, only this second peak is visible.
   \item in region {\em 4}, in some but not all of the observations (see
   table \ref{tab:zone}) a peak centered on a frequency between $\sim$6 and 7 Hz
   is detected.
   Its {\it rms} amplitude value is similar to that of second peak of region {\em
   3}, with the same energy dependence.
   \item in region {\em 5}, no QPOs are detected. 
\end{enumerate}

\begin{figure}
\begin{tabular}{c}
\resizebox{15cm}{!}{\includegraphics{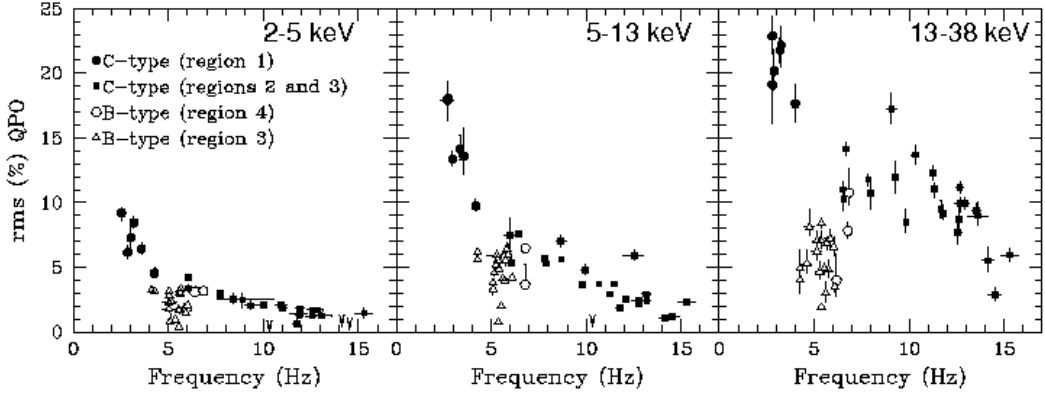}}
\end{tabular}
\caption{Fractional {\em rms} of the detected QPOs in all observations as function of 
their frequency for three energy bands. Error bars are estimated at 1 $\sigma$ confidence level.
Arrows indicate 3 $\sigma$ upper limits for type-C QPOs only. Filled symbols indicate type-C
QPOs, open symbols type-B.
Left panel: energy band 2-5 keV (PCA channels 0-13). Central panel: energy band 5-13 keV
(PCA channels 14-35). Right panel: energy band 13-38 keV (PCA channels 36-103).}
\label{fig:rms-all}
\end{figure}

\begin{figure}
\begin{tabular}{c}
\resizebox{15cm}{!}{\includegraphics{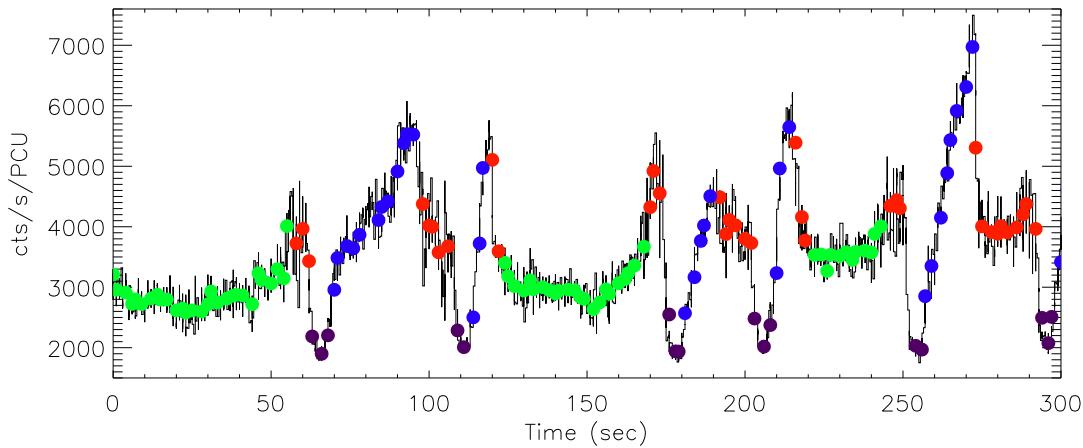}}
\end{tabular}
\caption{300-second light curve (2-13 keV energy band) for observation \# 5. Bin size is 0.125
sec. Every colored
point is averaged on 2 seconds. Different colors correspond to different regions in the
HHID in Figure 3: Green = region {\em 2}; Red = region {\em 3}, Purple =
region {\em 4}; Blue = region {\em 5}.}
\label{fig:licu}
\end{figure}

We will refer to the peak with centroid frequency ranging from $\sim$2 up to 15 Hz as 
{\it type-C} QPO
and to the peak with 4--7 Hz frequency detected in regions {\em 3} and
{\em 4} as {\it type-B} QPO. This identification will be motivated below.
A plot of the fractional {\it rms} amplitude
of all the detected peaks as a function of their centroid frequency, for each of three
energy bands analyzed, helps to clearly discriminate between the two types of QPO (see
Fig. \ref{fig:rms-all} ). The figures contain the data for all detections.
If for one observation a significant QPO
was detected only in one band, we repeated the fit fixing centroid and FWHM of the peak and
obtained its fractional {\it rms}.
Type-C QPOs are indicated with filled symbols, type-B QPOs with open symbols.
It is evident that the two QPOs correspond to different behaviours in these diagrams. Type-B
QPOs cluster in a small region of the diagrams, while type-C QPOs follow a trend.\\
The points with large errors in frequency correspond to detections where, due to the
changing frequency of the type-C QPO, we need two Lorentzian components to obtain a
satisfactory fit. For these points, we plot the average of the two centroid frequencies.\\
From figure \ref{fig:cd_hid_pds} we can relate the presence of the two QPO-types to
different zones of the source path in the HHID but we do not have any detailed information about
the time evolution of the source position in that diagram. We know that $\mu$ 
and $\beta$ class are characterized by fast variability and frequent spectral transitions
but we do not know precisely between which spectral
states the transitions happen and when. A light curve where parts corresponding to
different regions in the HHID are colored with the same colors used in Figure
\ref{fig:cd_hid_pds} can help us to clarify this point (see Fig \ref{fig:licu}).
We chose a 300-second interval representative of the source behaviour. As  
mentioned above, we do not consider region {\em 1} for this analysis. If, using the
state classification presented in \cite{2000A&A...355..271B}, we make the identifications
Green = C-state, Purple = A-state, Blue = B-state and we consider the Red as a
Transition (T) state, we have the following state sequences: CTAB TAB BTC.

\section{Conclusions}
We analyzed five RXTE/PCA observations and detected in all of them peaks that we identify with two
different QPO-types: type-C and type-B QPO. This is the first detection of a type-B QPO in GRS
1915+105. In all the five observations we detected the type-B together with the type-C,
in two of them we also detected the type-B alone (see Tab. \ref{tab:zone}). We could not
correlate the presence of the type-B in the fourth (Purple) region of Fig. \ref{fig:cd_hid_pds} neither
with the selected variability class nor with the count rate or the spectral hardness.\\
To date,  a type-B QPO has been detected in few systems, associated to spectral transitions from
HIMS to SIMS. In GRS 1915+105 is well known the association between the X-ray and the radio
activity (see \cite{1997MNRAS.292..925P}, \cite{1998A&A...330L...9M} and \cite{2004ARA&A..42..317F}
for a review), therefore the non-detection of a type-B QPO was rather interesting, as alreday pointed
out in \cite{2005A&A...440..207B}. Our detection of the type-B QPO in GRS 1915+105 strenghtens two
points:
\begin{enumerate}
 \item GRS 1915+105 behaves like all other radio-loud BHCs: the type-B appears
 in correspondence of spectral transition from the B-state to the A-state (Transition, Red) and when
 the source is
 in the A-state (Purple). In the light curve in Fig. \ref{fig:licu} we see state oscillations 
 CTAB TAB TAB that we can identify as fast passages between the HIMS and the SIMS
 of the other BHCs (\cite{2004MNRAS.355.1105F}, \cite{2004A&A...426..587C}). The
 type-B appears also in correspondence of the transition from the B-state to the C-state
 (Blue-Red-Green in Fig. \ref{fig:licu}), so not in an oscillation event but in a transition from a
 soft to a hard state.
 \item the association between type-B QPO, spectral transitions and radio jets emission. If we consider
 the type-B QPO as the signature radio jet  emission, we have this ``signature'' also in the
 prototypical galactic jet source. 
\end{enumerate} 

Klein-Wolt et al. (2002, see \cite{2002MNRAS.331..745K}) compared 101 simultaneous Ryle Telescope
radio and RXTE X-ray observations of GRS 1915+105: they found that $\beta$ class
observations are associated with strong radio oscillations while $\mu$ observations are positively
associated to weak radio activity without oscillations. Unluckily, they did not analyze any of the
observations from our data set so we can not verify directly GRS 1915+105 radio behaviour while
the type-B QPO was present.

\end{document}